\documentclass[aps,prb,reprint,superscriptaddress,10pt,longbibliography,nofootinbib,floatfix]{revtex4-2}
\usepackage{amsmath,amssymb,graphicx,bm,booktabs,microtype,array}
\usepackage{float}
\usepackage[colorlinks=true,citecolor=blue,urlcolor=blue,linkcolor=blue]{hyperref}
\newcommand{\Tr}{\operatorname{tr}}
\begin{document}
\raggedbottom
\title{Critical--localized reentry under periodic spacer control}
\author{Xin Wang}
\thanks{Xin Wang and Pei-Jie Chang contributed equally to this work.}
\affiliation{School of Physical Science and Technology, Nantong University, Nantong, 226019, China}

\author{Pei-Jie Chang}
\thanks{Xin Wang and Pei-Jie Chang contributed equally to this work.}
\affiliation{State Key Laboratory of Low-Dimensional Quantum Physics and Department of Physics, Tsinghua University, Beijing 100084, China}

\author{Dong Ruan}
\email{dongruan@tsinghua.edu.cn}
\affiliation{State Key Laboratory of Low-Dimensional Quantum Physics and Department of Physics, Tsinghua University, Beijing 100084, China}
\affiliation{Frontier Science Center for Quantum Information, Tsinghua University, Beijing 100084, China}

\author{Gui-Lu Long}
\email{gllong@tsinghua.edu.cn}
\affiliation{State Key Laboratory of Low-Dimensional Quantum Physics and Department of Physics, Tsinghua University, Beijing 100084, China}
\affiliation{Frontier Science Center for Quantum Information, Tsinghua University, Beijing 100084, China}
\affiliation{Beijing Academy of Quantum Information Sciences, Beijing 100193, China}
\affiliation{Beijing National Research Center for Information Science and Technology, Beijing 100084, China}
\begin{abstract}
Tracking a fixed fraction of a quasiperiodic spectrum tests whether repeated critical--localized alternation extends over a nonzero spectral weight. We find such alternation in a matched-dimer chain by varying only the spacer on-site energies and following a fixed interval of integrated-density-of-states (IDS) labels. An exact Bloch-determinant identity links the physical transfer discriminant to the bands of a periodic reference operator. For the golden cell frequency, we prove that, on a common full-measure set of phases and for every finite value of the spacer control, the physical spectrum in the open reference bands is critical, with purely singular-continuous spectrum and zero Lyapunov exponent, while the spectrum strictly outside the bands is pure point, with exponentially decaying eigenfunctions. Validated IDS bounds certify repeated critical--localized alternation of the same fixed spectral fraction at successive spacer controls. Spacer shifts move the physical levels and reference-band edges at different rates, allowing us to reshape finite-chain return windows and place their endpoints at prescribed controls. An affine recalibration relocates the control family exactly.
\end{abstract}
\maketitle

\section{Introduction}
Anderson showed that random on-site energies can suppress diffusion in a lattice \cite{Anderson1958}. Scaling theory relates localization to the size dependence of conductance \cite{Abrahams1979}. The Aubry--Andr\'e--Harper (AAH) model exhibits an extended--localized transition driven by an incommensurate potential \cite{Harper1955,AubryAndre1980,Hofstadter1976,Jitomirskaya1999}. Generalizations admit mobility edges separating extended and localized states \cite{DasSarma1988,Biddle2010,Biddle2011,Ganeshan2015,WangMosaic2020,Vu2023}. Quasiperiodic lattices also support critical states with hierarchical spatial profiles, as in Fibonacci chains \cite{Kohmoto1983,Ostlund1983,Sutherland1987,Kohmoto1987}, and fractal or multifractal behavior with shallow potentials or power-law hopping \cite{Yao2019,Deng2019}. Experiments have observed the AAH transition in ultracold atoms and photonic lattices \cite{Roati2008,Lahini2009}. Critical phases have been proposed for optical Raman lattices \cite{WangCritical2020} and probed in superconducting circuits \cite{Huang2026}.

Competing lattice scales can replace a single localization transition with reentry, as in dimerized chains \cite{Roy2021,Roy2022,WangLongRangeReentry2023}, off-diagonal Aubry--Andr\'e--Fibonacci chains \cite{Tabanelli2024,Yang2026}, and lattices with competing periodic and quasiperiodic potentials \cite{Chang2026}. The periodic field itself can serve as the control while the quasiperiodic modulation remains fixed \cite{Padhan2022,Qi2023}. Reentry can also involve critical states: Allard and Weick found critical levels between localized regimes in a quasiperiodic dipolar chain \cite{AllardWeick2026}, while Wei et al.\ tracked repeated entry into the critical sector at a fixed normalized rank in a quasiperiodic Raman lattice \cite{Wei2026}. This motivates following a fixed interval of integrated-density-of-states (IDS) labels, which tracks a nonzero spectral weight as its energy endpoints move. Can this entire fraction alternate between critical and localized sectors at successive controls?

Understanding this alternation requires locating the spectral boundaries crossed by the selected fraction. Anomalous mobility edges separate critical and localized spectra \cite{Liu2022,WangCoexist2022,Duncan2024,Banerjee2025}. Matched-dimer mosaics combine zero-energy antisymmetric orbitals with arbitrarily weak quasiperiodic bonds \cite{Zhou2023}, and singular Jacobi analysis distinguishes their spectral types \cite{WangZhou2025}. Periodic reference spectra organize localization in generalized mosaics \cite{LonghiMosaic2024}. Sector assignment depends on the motion of the physical spectrum relative to these boundaries \cite{WangEngineering2023}. This suggests using periodic on-site control at fixed quasiperiodic bonds to produce complete returns of the selected fraction and raises a further question: can the entry and exit controls of a finite-chain return be tuned independently?

We address these questions in a spacer-decorated matched-dimer chain with fixed quasiperiodic bonds and tunable spacer on-site energies. Compression onto the antisymmetric dimer orbitals and spacers gives a periodic reference operator whose open bands coincide exactly with the physical sector $|\Delta|<2$, where $\Delta$ is the physical transfer discriminant. For the golden cell frequency, we prove on a common full-measure phase set for every finite $g$ that this sector is critical (C), with purely singular-continuous spectrum and zero Lyapunov exponent, and that the exterior $|\Delta|>2$ is exponentially localized (L) with pure-point spectrum. Validated IDS bounds certify the C--L--C--L--C--L sequence for the same interval, and real-space profiles show how its spatial support changes across the returns. In the third finite-chain return, the highest selected level reaches the upper reference-band edge as the window closes. Because the level and the edge have different spacer weights, an offset can move this crossing and widen or suppress the return. Two independent offsets allow us to place the finite-chain endpoints at prescribed controls.

\begin{figure}[!tb]
\centering
\includegraphics[width=\columnwidth]{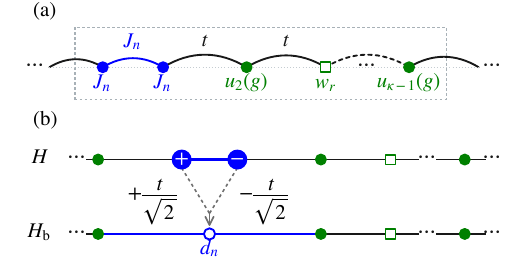}
\caption{Decorated quasiperiodic chain and its periodic compression. (a) A cell with a matched dimer (blue), whose hopping and on-site energies equal $J_n$, and spacers (green). The open square marks a spacer with a control-independent energy $w_r$. (b) Compression of the physical chain $H$ onto its zero-energy dark orbitals $d_n$ and spacers gives the periodic reference chain $H_{\rm b}$ with contacts $\pm t/\sqrt2$, where $t$ is the regular hopping.}
\label{fig:mechanism}
\end{figure}

\section{Model and fixed-weight spectral protocol}\label{sec:model}
We study a decorated chain with quasiperiodic dimer bonds and a static control $g$ that acts only on the periodic spacer on-site energies.
Each cell $n\in\mathbb Z$ contains a matched dimer at sites $r=0,1$ and $\kappa-2$ spacers at $r=2,\ldots,\kappa-1$, with $\kappa\geq3$ [Fig.~\ref{fig:mechanism}(a)]. The dimer hopping and both dimer on-site energies equal $J_n$. Taking the regular hopping $t$ as the energy unit ($t=1$ below), we write
\begin{align}
H(g)={}&\sum_n J_n\bigl(c^\dagger_{n,0}c_{n,1}+\mathrm{H.c.}
+n_{n,0}+n_{n,1}\bigr)\nonumber\\
&+\sum_n\sum_{r=1}^{\kappa-2}\bigl(c^\dagger_{n,r}c_{n,r+1}+\mathrm{H.c.}\bigr)\nonumber\\
&+\sum_n\bigl(c^\dagger_{n,\kappa-1}c_{n+1,0}+\mathrm{H.c.}\bigr)
+\sum_n\sum_{r=2}^{\kappa-1}u_r(g)n_{n,r},\label{eq:H}
\end{align}
with $J_n=2\cos(2\pi\alpha n+\theta)$ and $u_r(g)=w_r+v_rg$.
Here $c_{n,r}$ annihilates a particle at site $r$ of cell $n$, $n_{n,r}=c^\dagger_{n,r}c_{n,r}$, and $\mathrm{H.c.}$ denotes the Hermitian conjugate. The quasiperiodic frequency per cell is $\alpha=(\sqrt5-1)/2$, and $\theta$ is a phase in radians. The offsets $w_r$ are in units of $t$, and the slopes $v_r$ are dimensionless. The Hamiltonian depends linearly on $g$,
\begin{equation}
H(g)=H(0)+gW,\qquad W=\sum_n\sum_{r=2}^{\kappa-1}v_rn_{n,r}.\label{eq:affine}
\end{equation}

We track the same fraction of the spectrum as $g$ varies. In an open chain with $L$ sites, the normalized interval $I=[f_-,f_+]$, with $0<f_-<f_+<1$, selects the inclusive set of zero-based ranks
\begin{equation}
I_L=\{\lfloor f_-L\rfloor,\ldots,\lfloor f_+L\rfloor\}.\label{eq:rank}
\end{equation}
In the infinite chain, $f$ labels the energy quantile $E_g(f)=\inf\{E:\mathcal N_g(E)\ge f\}$ of the IDS per physical site \cite{Ostlund1984}. The energy endpoints vary with $g$ at fixed IDS weight $\mathcal W=f_+-f_-$. The finite-chain implementation in Eq.~\eqref{eq:rank} contains $\lfloor f_+L\rfloor-\lfloor f_-L\rfloor+1=\mathcal W L+O(1)$ ranks. Each chain is ordered separately at its own control and phase. Alternation of a fixed IDS interval requires all its energy quantiles to share the same spectral type at each tested control.

\section{Periodic reference bands and analytical spectral classification}\label{sec:transfer}
\subsection{Dark-orbital compression and reference bands}\label{sec:compression}
Compression onto the zero-energy antisymmetric dimer orbitals (dark orbitals) and the spacers produces a periodic reference Hamiltonian.
For every $J_n$, the isolated matched dimer has a zero-energy antisymmetric orbital $|d_n\rangle=(|n,0\rangle-|n,1\rangle)/\sqrt2$, where $|n,i\rangle$ denotes the state with one particle occupying site $i$ of cell $n$. Let $\mathcal P_d$ project onto these dark orbitals and all spacer sites. The Hamiltonian projected onto this subspace is
\begin{equation}
H_{\rm b}=\left.\mathcal P_dH\mathcal P_d\right|_{\operatorname{Ran}\mathcal P_d}.\label{eq:compression}
\end{equation}
The antisymmetric combination cancels the quasiperiodic dimer terms exactly, leaving a zero-energy orbital that is identical in every cell; together with the periodically controlled spacers, this makes the compressed chain periodic, with contacts $\pm1/\sqrt2$ [Fig.~\ref{fig:mechanism}(b)]. In the full chain the symmetric orbitals still couple to the retained sector, so $H_{\rm b}$ supplies auxiliary reference bands and $H$ governs the physical eigenstates.

The periodic reference bands are obtained from the Bloch matrix $h_{\rm b}(K;g)$ of the compressed chain. Each cell retains one dark orbital and $\kappa-2$ spacers, so this matrix has dimension $(\kappa-1)\times(\kappa-1)$. Here $K\in[0,2\pi)$ is the Bloch phase acquired across one cell. At fixed $g$, its eigenvalues trace out the reference bands as $K$ varies. We index these bands in ascending energy and denote their lower and upper edges by $B_j^-(g)$ and $B_j^+(g)$.

\subsection{Transfer reduction and the band identity}\label{sec:band-identity}
The physical transfer discriminant links propagation in the quasiperiodic chain to the periodic reference bands.
Let $\psi_{n,r}$ denote the amplitude at site $r$ of cell $n$ and collect the amplitudes on the two sides of its left boundary in $X_n=(\psi_{n,0},\psi_{n-1,\kappa-1})^{\mathsf T}$. For $J_n\ne0$, the cell transfer matrix $M_n$ propagates this vector across cell $n$ according to $X_{n+1}=M_n X_n$. Matching the dimer hopping and both on-site energies cancels the terms proportional to $J_n$ in the dimer transfer matrix and gives
\begin{equation}
M_n=A+\frac{B}{J_n},\qquad \det A=1,\quad \operatorname{rank}B=1.\label{eq:M}
\end{equation}
The matrices $A(E,g)$ and $B(E,g)$ are phase independent. Their traces are
\begin{equation}
\Delta(E,g)=\Tr A(E,g),\qquad \tau(E,g)=\Tr B(E,g).\label{eq:delta}
\end{equation}
An exact change of coordinates takes $X_n$ to $(y_n,-y_{n-1})^{\mathsf T}$ and is invertible for $\tau\ne0$, including throughout $|\Delta|<2$ (Appendix~\ref{app:transfer}). The scalar coordinate $y_n$ obeys
\begin{equation}
y_{n+1}+y_{n-1}-\frac{\tau(E,g)}{2\cos(2\pi\alpha n+\theta)}y_n
=\Delta(E,g)y_n.\label{eq:sec}
\end{equation}
The same discriminant also determines the periodic reference bands introduced in Sec.~\ref{sec:compression} through the exact identity
\begin{equation}
2\det[E-h_{\rm b}(K;g)]=2\cos K-\Delta(E,g).\label{eq:backbone}
\end{equation}
The equation $2\cos K=\Delta$ has a real solution exactly when $|\Delta|\leq2$. The open band interiors satisfy $|\Delta|<2$, and the band edges satisfy $|\Delta|=2$. Equation~\eqref{eq:backbone} holds also at spacer eigenenergies (Appendix~\ref{app:transfer}).

For a fixed IDS label $f$, the band identity locates the energy quantile $E_g(f)$ relative to the reference bands through $\Delta(E_g(f),g)$. As $g$ varies, both the quantile energy and the reference bands move. The linear dependence of $H(g)$ on $g$ does not require $\Delta(E_g(f),g)$ to be monotonic, so repeated crossings of the band boundaries $|\Delta|=2$ are possible.

\subsection{Spectral classification of the open sectors}\label{sec:spectral}
Sections~\ref{sec:compression} and~\ref{sec:band-identity} locate the transition boundaries but do not determine the phases on either side.

The irrational modulation produces arbitrarily weak bonds in both directions. Cutting a sequence of weak bonds whose total hopping strength is finite separates the chain into finite segments. This changes the Hamiltonian by a trace-class perturbation, which preserves its absolutely continuous part by the Kato--Rosenblum theorem. The disconnected finite segments have no absolutely continuous spectrum, so neither does the original chain, for every phase and finite $g$. This rules out a conventional extended sector. In the open bands, a Fourier argument applied to Eq.~\eqref{eq:sec} also excludes normalizable eigenstates for the nonresonant phases specified in Appendix~\ref{app:proof} \cite{WangZhou2025}, leaving purely singular-continuous spectrum. In the exterior, an exact fixed-energy reduction to the matched-dimer parent model gives exponentially decaying Green functions on suitable finite blocks. These estimates, combined with the inverse transformation and spectral expansion, establish pure-point spectrum with exponentially decaying physical eigenfunctions.

On the physical spectrum $\Sigma_g$ of $H(g)$, the discriminant gives the phase-averaged Lyapunov exponent $\gamma(E,g)$ per physical site (Appendix~\ref{app:lyapunov}),
\begin{equation}
\gamma(E,g)=\frac1\kappa
\begin{cases}
0,&|\Delta(E,g)|\le2,\\
\operatorname{arcosh}(|\Delta(E,g)|/2),&|\Delta(E,g)|>2.
\end{cases}\label{eq:exactLE}
\end{equation}
Equation~\eqref{eq:exactLE} gives a zero Lyapunov exponent also at spectral energies on a band edge. For physical spectral energies in the open reference bands, pure singular continuity together with the zero Lyapunov exponent identifies the critical sector, consistent with the incommensurate-zero framework \cite{ZhouFramework2026}.

For the golden cell frequency, the following classification holds on a common full-measure phase set for every finite $g$. The physical spectrum in the open reference bands, $|\Delta(E,g)|<2$, is critical, with purely singular-continuous spectrum (Appendix~\ref{app:proof}). Strictly outside the reference bands, $|\Delta(E,g)|>2$, it is localized, with pure-point spectrum and exponentially decaying eigenfunctions (Appendix~\ref{app:localization}). Combined with the IDS bounds below, this classification allows us to certify repeated critical--localized alternation of the same nonzero spectral fraction under periodic spacer control at fixed quasiperiodic bonds.

\section{Fixed-weight reentry: IDS certification and spatial signatures}\label{sec:results}
The nine-site example ($\kappa=9$) uses the fixed IDS interval $I=[0.558,0.571]$. A finite-chain search selected the spacer parameters and rank window (Appendix~\ref{app:selection}). The slopes are $\bm v=(0,0,0,\tfrac12,1,0,1,-1,1)$, and the rounded offsets are $\bm w=(0,\allowbreak 0,\allowbreak 3.4,\allowbreak -6.13,\allowbreak -0.94,\allowbreak -6.05,\allowbreak -7.05,\allowbreak 0.28,\allowbreak -3.62)$.
These parameters and the IDS interval remain fixed in the baseline control scan and size comparisons without additional spacer offsets. Unless otherwise stated, the finite-chain calculations use $L=14373$ sites and $\theta=0.37$.

\subsection{Band geometry and complete-window returns}\label{sec:complete-windows}
A complete-window return requires every selected level to lie inside the open reference bands.
Let $E_{m,L}(g,\theta)$ be the open-chain eigenvalues in ascending order, indexed from $m=0$. To identify the controls satisfying this condition for $I_L$, we define
\begin{equation}
G_{I,L}(g,\theta)=\max_{m\in I_L}|\Delta(E_{m,L}(g,\theta),g)|-2.\label{eq:window-G}
\end{equation}
Write $G=G_{I,L}(g,\theta)$ when the chain and phase are clear. A complete-window return has $G<0$. At $G=0$ at least one selected level lies at a band edge, and at $G>0$ at least one lies in the exterior. To test whether all selected levels lie strictly outside the reference bands, we also define $F_{\min}=\min_{m\in I_L}|\Delta(E_{m,L},g)|-2$; the entire rank window is in the exterior when $F_{\min}>0$.

\begin{figure}[!tb]
\centering
\includegraphics[width=\columnwidth]{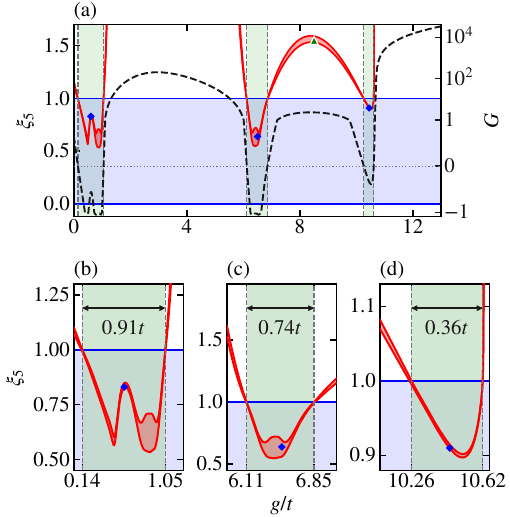}
\caption{Band geometry of $I$ at $L=14373$. (a) Blue shading spans reference band 5 in coordinates $\xi_5=(E-B_5^-)/(B_5^+-B_5^-)$. Red shading spans the energy envelope of the 187 selected physical levels. The dashed black curve shows $G$ on the right symmetric-log axis (linear for $|G|\le1$). (b)--(d) Local returns near $C_1,C_2,C_3$, with different axis ranges and horizontal-axis ticks at the return endpoints. Double arrows give the return widths. Diamonds mark critical controls, and the green triangle in (a) marks the localized control $L_2$. Green shading and vertical dashed lines indicate $G<0$ and $G=0$, respectively.}
\label{fig:backbone}
\end{figure}

Figure~\ref{fig:backbone} resolves the 187-level energy envelope as it moves relative to band 5. At the three controls $C_1,C_2,C_3$ with $g/t=0.6,6.5,10.45$, the entire envelope lies inside band 5; at the boundaries of a return, the envelope reaches a band edge and the window ceases to be complete. Complete containment in a single band $j$ requires
\begin{equation}
B_j^-(g)<\min_{m\in I_L}E_{m,L}\leq\max_{m\in I_L}E_{m,L}<B_j^+(g).
\label{eq:window-containment}
\end{equation}
The subsequent exterior controls $L_1,L_2,L_3$ are at $g/t=3.5,8.5,12$. The same rank window must lie entirely inside the reference bands at each $C_i$ and entirely outside them at each $L_i$. This simultaneous requirement constrains both the position and the weight of the selected window. At fixed weight $\mathcal W=f_+-f_-$, varying $f_-$ shifts the window through the spectrum. Increasing $\mathcal W$ includes additional levels, which may violate the required sector condition at one or more controls. The admissible range of $f_-$ therefore narrows as the window weight increases, as observed in the finite-chain scan (Appendix~\ref{app:finite-size}).

For $L=14373$, intersecting the intervals of complete band containment over phases $0.17,1.31,2.73$ gives three approximate intervals of $g/t$: $(0.14,1.05)$, $(6.11,6.85)$, and $(10.26,10.62)$. These intervals agree to the displayed precision with Fig.~\ref{fig:backbone}(b)--(d). Recomputing all six endpoints at the larger sizes $L=37629$ and $60885$ for the same phases gives the same phase-intersection intervals to the displayed precision, supporting the stability of the return-window positions and widths over the sampled sizes and phases. Appendix~\ref{app:control-checks} reports the corresponding checks over different sizes and phases.

\subsection{IDS certification of the fixed spectral fraction}\label{sec:certification}
The finite-chain envelopes locate three complete returns, but the corresponding nonzero spectral weight must be established from infinite-chain IDS bounds. We therefore determine a common IDS interval whose energies lie entirely in alternating critical and localized sectors across the six control neighborhoods. For an energy enclosure $[a,b]$, the required inequalities are $\mathcal N_g(a)<f_-$ and $\mathcal N_g(b)>f_+$. Monotonicity of the IDS then places every $E_g(f)$ with $f\in I$ inside $[a,b]$.

Let $n_L(E,\phi)$ count eigenvalues strictly below $E$ in an open block of $L$ sites containing an integer number of cells. Cutting the bond between adjacent blocks changes the count by at most one per interface, because the removed hopping has one positive and one negative eigenvalue. The IDS is continuous (Appendix~\ref{app:IDS}). Dividing the count estimate by the number of sites and taking the ergodic limit gives
\begin{equation}
\left|\mathcal N_g(E)-\frac1L\int_0^{2\pi}n_L(E,\phi)\frac{\mathrm d\phi}{2\pi}\right|\leq\frac1L.\label{eq:IDSblock}
\end{equation}
To bound the phase integral, use $K$ uniformly spaced phases $\phi_j$. Every phase is within $\pi/K$ of a sample, and $\|H(\phi)-H(\phi')\|\le4|\phi-\phi'|$. Choosing $\epsilon_\uparrow\ge4\pi/K$ therefore bounds the corresponding eigenvalue shifts. Eigenvalue monotonicity yields
\begin{equation}
\begin{gathered}
\frac1{KL}\sum_jn_L(E-\epsilon_\uparrow,\phi_j)-\frac1L\leq\mathcal N_g(E),\\
\mathcal N_g(E)\leq\frac1{KL}\sum_jn_L(E+\epsilon_\uparrow,\phi_j)+\frac1L.
\end{gathered}\label{eq:IDSgrid}
\end{equation}
The validated IDS bounds identify label ranges whose energy quantiles remain within the certified energy enclosures throughout each control neighborhood. Table~\ref{tab:occupation} lists these ranges and their spectral types; the underlying energy enclosures and IDS bounds are given in Table~\ref{tab:certificates} of Appendix~\ref{app:IDS}.

\begin{table}[tb]
\caption{Certified IDS ranges and spectral types in the six control neighborhoods $|g-g_i|\le5\times10^{-6}t$. The label ranges are rounded inward, and the discriminant bounds hold over the corresponding full energy--control rectangles.}
\label{tab:occupation}
\setlength{\tabcolsep}{4pt}
\begin{ruledtabular}
\begin{tabular}{cccc}
$g_i/t$ & IDS labels & Discriminant bound & Type \\
\hline
$0.6$ & $(0.5368,0.5752)$ & $|\Delta|\le1.83$ & C \\
$3.5$ & $(0.5563,0.5793)$ & $\Delta\ge92.8$ & L \\
$6.5$ & $(0.5144,0.5923)$ & $|\Delta|\le1.82$ & C \\
$8.5$ & $(0.5577,0.5915)$ & $\Delta\ge2.12$ & L \\
$10.45$ & $(0.5569,0.5804)$ & $|\Delta|\le1.98$ & C \\
$12$ & $(0.5569,0.5846)$ & $\Delta\ge1.26\times10^4$ & L \\
\end{tabular}
\end{ruledtabular}
\end{table}

Intersecting the six label ranges in Table~\ref{tab:occupation} gives the common IDS range $\mathcal J=(0.5577,0.5752)$, with width $|\mathcal J|=0.0175$.
For every closed label interval strictly inside $\mathcal J$, the associated energies lie within the corresponding certified energy enclosure at each of the six control neighborhoods.

The discriminant bounds in Table~\ref{tab:occupation} place these enclosures strictly inside the reference bands ($|\Delta|<2$) for the C neighborhoods and strictly outside them ($\Delta>2$) for the L neighborhoods. Applying the spectral classification in Sec.~\ref{sec:spectral} certifies C--L--C--L--C--L for every such label interval. Thus, the certified alternating spectral weight can be made arbitrarily close to $0.0175$ from below.

The fixed interval $I=[0.558,0.571]$ used here lies strictly inside $\mathcal J$, so its entire IDS weight of $0.013$ has this certified alternation. Appendix~\ref{app:control-checks} reports the corresponding finite-chain checks over multiple sizes and phases.

\subsection{Real-space support and level spacings}
\begin{figure}[!tb]
\centering
\includegraphics[width=\columnwidth]{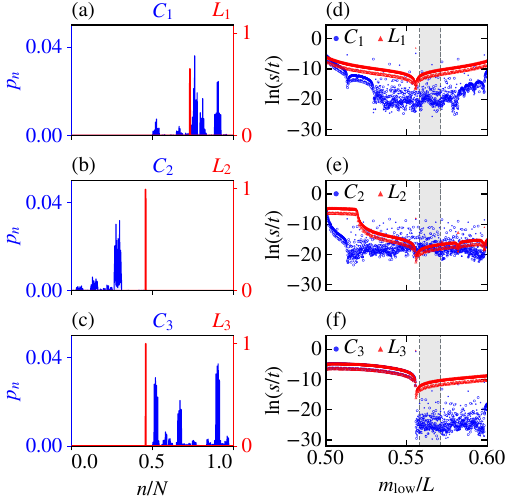}
\caption{Real-space probability distributions and adjacent level spacings for the three returns at $L=14373$. Each row corresponds to one critical control $C_i$ and its subsequent localized control $L_i$. (a)--(c) Cell probabilities $p_n=\sum_r|\psi_{n,r}|^2$ for the open-chain eigenstate at zero-based rank $m=8113$. Blue curves correspond to $C_i$ (left axes), and red curves to $L_i$ (right axes). (d)--(f) Logarithmic spacings $\ln(s/t)$ for a periodic chain. Filled and open markers denote $s_{\rm eo}$ and $s_{\rm oe}$, respectively. Circles (blue) denote $C_i$ and triangles (red) denote $L_i$. Dashed lines delimit $I=[0.558,0.571]$. Gaps at or below the roundoff threshold are omitted; all gaps within $I$ are retained.}
\label{fig:realspace}
\end{figure}

At the certified critical controls $C_i$, the representative finite-chain states spread over several separated clusters with strongly uneven weights [Fig.~\ref{fig:realspace}(a)--(c)]. The locations and relative weights of these clusters change between returns. Thus, reentry into the critical sector does not restore the same spatial distribution. At the localized controls $L_i$, the states at the same rank and phase concentrate most of their weight on one or two cells. Their maximum cell probabilities are $0.65$, $0.99$, and $1.00$ for $L_1,L_2,L_3$, respectively.

The dimer hopping varies quasiperiodically and takes arbitrarily small magnitudes in the infinite chain, so propagation encounters weak links on many length scales. This clustered support is consistent with the analytical classification of the physical spectrum inside the open reference bands as critical (Sec.~\ref{sec:spectral}).

The contrast also appears when we compare the two alternating sequences of adjacent level spacings under periodic boundary conditions. For eigenvalues ordered using one-based indices, these sequences are $s_{\rm eo}=E_{2j}-E_{2j-1}$ and $s_{\rm oe}=E_{2j+1}-E_{2j}$. The horizontal coordinate is $m_{\rm low}/L$, where $m_{\rm low}$ is the lower one-based rank of each pair. Within the shaded rank interval, both sequences at $C_i$ form irregular clouds across a broad range of spacings, while those at the paired $L_i$ trace smoother branches [Fig.~\ref{fig:realspace}(d)--(f)]. In the second pair, the clouds and branches partly overlap in magnitude, but their variation with rank is distinct. In the third pair, the spacings at $C_3$ also reach much smaller values than those at $L_3$. The irregular spacing sequences complement the clustered profiles at these critical controls \cite{Banerjee2025}.

\subsection{Inverse participation ratio and size dependence}
\begin{figure}[!t]
\centering
\includegraphics[width=\columnwidth]{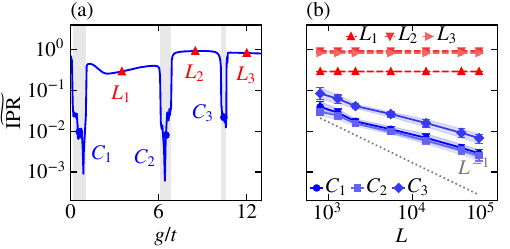}
\caption{$\widetilde{\mathrm{IPR}}$ for the fixed IDS interval $I$. (a) Control scan at $L=14373$. Circles and triangles mark C and L controls, respectively. Shading marks the controls at which all selected energies lie in open reference bands. (b) Size dependence over all selected ranks and twelve uniformly spaced phases at seven sizes from $L=801$ to $60885$. Shading and capped bars span the phase-specific geometric means; the dotted guide scales as $L^{-1}$. Both panels share the vertical scale.}
\label{fig:scan}
\end{figure}

The inverse participation ratio (IPR) extends the comparison from representative profiles to all selected ranks and tests how their spatial support changes with chain size.
For a normalized eigenvector, let $p_j=|\psi_j|^2$ be the probability at physical site $j$. The IPR and its geometric mean are \cite{EversMirlin2008,Rodriguez2011}
\begin{equation}
\mathrm{IPR}=\sum_{j=0}^{L-1}p_j^2,\qquad
\widetilde{\mathrm{IPR}}=\exp\langle\ln\mathrm{IPR}\rangle.\label{eq:IPR}
\end{equation}
The reciprocal $1/\mathrm{IPR}$ is the effective number of occupied sites. The tilde denotes a geometric mean, with equal weight assigned to every selected rank and, for size comparisons, to each sampled phase. The size comparisons in Figs.~\ref{fig:scan}(b), \ref{fig:control}(d), and \ref{fig:inverse-design}(b) use the same twelve uniformly spaced phases over a full phase period at every size. At fixed $g$, we fit $\ln\widetilde{\mathrm{IPR}}$ against $\ln L$,
\begin{equation}
\ln\widetilde{\mathrm{IPR}}(L)\simeq c-D_2^{\rm fit}\ln L,
\label{eq:D2}
\end{equation}
where $c$ is the fitted intercept and $D_2^{\rm fit}$ is the effective scaling exponent over the sampled sizes.

For a conventional extended state, the weight is of order $1/L$ on order $L$ sites, giving $\mathrm{IPR}\sim L^{-1}$. For an exponentially localized state, the IPR approaches a size-independent value once $L$ exceeds the localization length. Critical states can exhibit intermediate scaling, $\mathrm{IPR}\sim L^{-D_2}$ with $0<D_2<1$, reflecting spatial support that grows with $L$ but remains strongly nonuniform. The finite-size fits characterize this behavior over the sampled range.

Figure~\ref{fig:scan}(a) shows the typical IPR over all selected ranks $I_L$ as $g$ varies. The three low-IPR intervals align with the shaded complete-window returns determined by the finite-chain band-containment criterion of Sec.~\ref{sec:complete-windows}.

The size dependence in Fig.~\ref{fig:scan}(b) distinguishes these regions. At $L_i$, adding sites leaves the typical IPR nearly constant, as expected when the wave-function weight remains concentrated on a finite number of sites. At $C_i$, the typical IPR decreases as the chain grows, indicating that the effective number of sites carrying weight increases across the sampled sizes. The decrease is slower than the $L^{-1}$ guide for a uniformly extended state, with pooled effective slopes $D_2^{\rm fit}\simeq0.58,0.57,0.57$ for $C_1,C_2,C_3$, respectively, over the seven sizes (Appendix~\ref{app:finite-size}). This intermediate scaling is consistent with the critical classification and the clustered profiles in Fig.~\ref{fig:realspace}. The typical IPR at $C_3$ stays above those at $C_1$ and $C_2$ over the sampled sizes. The representative $C_3$ profile likewise concentrates more weight in its largest peaks [Fig.~\ref{fig:realspace}(c)].

\section{Microscopic control and inverse design of return windows}\label{sec:control}
The preceding sections establish repeated critical--localized returns for a fixed spacer configuration. Controlling the positions and widths of these return windows requires understanding how spacer offsets move the physical spectrum relative to the reference-band edges.

The third complete-window return ends at $g/t\simeq10.615$, where the highest selected level reaches the upper edge $B_5^+$ [Fig.~\ref{fig:backbone}(d)]. The bare spacer crossing $u_2=u_6$ occurs at $g/t=10.45$, and the gap between reference bands 5 and 6 reaches its minimum at $g/t\simeq10.620$. An offset $p$ on spacer site $r=3$ changes $w_3\mapsto w_3+p$ while preserving all scan slopes and the bare detuning $u_2-u_6=10.45t-g$.
We select spacer site $r=3$ for its large width response among the intervening sites (Appendix~\ref{app:control-checks}) and examine $-4\le p/t\le1$.

\subsection{Unequal spacer weights and window deformation}
\begin{figure}[!tbp]
\centering
\includegraphics[width=\columnwidth]{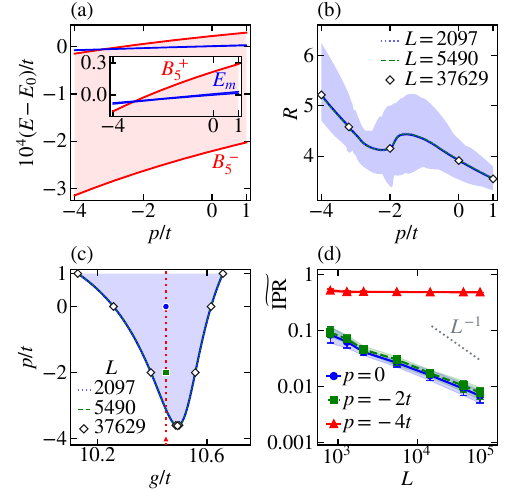}
\caption{Spacer control $w_3\mapsto w_3+p$. The main data in (a)--(c) use $L=14373$; (a), (b), and (d) use $g/t=10.45$. (a) Physical energy envelope (blue) and reference band (red), displayed as $10^4(E-E_0)/t$ with $E_0=3.5734t$. The inset magnifies the upper-edge crossing. (b) Spacer-weight ratio $R=b_3/\langle q_3\rangle_I$; shading spans the individual ratios $b_3/q_{3,m}$ over the selected ranks. (c) Blue shading marks complete returns ($G<0$), bounded by solid curves at $L=14373$; white includes partial or no returns. Open diamonds mark return boundaries and the termination point. The red dotted line marks $g/t=10.45$. (d) $\widetilde{\mathrm{IPR}}$ at seven sizes from $L=801$ to $60885$, using the twelve phases of Fig.~\ref{fig:scan}(b). Shading and capped bars span the phase-specific geometric means; the dotted guide scales as $L^{-1}$. The offset markers also appear in (c).}
\label{fig:control}
\end{figure}

At fixed $g/t=10.45$, Fig.~\ref{fig:control}(a) shows how the selected energy envelope and reference band respond to the spacer offset $p$. The blue envelope lies inside the red reference band at $p=0$. As $p$ decreases, the upper edge $B_5^+$ moves downward faster than the physical levels and passes through the envelope, as resolved in the inset. By $p=-4t$, the envelope lies above $B_5^+$. The lower edge $B_5^-$ remains below the envelope throughout, so the loss of containment occurs through the upper edge while the reference band stays open.

This unequal motion changes band containment at fixed $g$. To determine how it shifts the return-window endpoints, we compare the responses of the physical level and reference-band edge at their crossing.
Let $g_*(p)$ be a window endpoint at which only one selected level $E_{m_*}(g,p)$, called the active level, reaches the reference-band edge $B_{\rm edge}(g,p)$. Assume that the active level and band edge are simple, that the active rank is locally unchanged, and that the crossing is transverse, $\partial_gE_{m_*}-\partial_gB_{\rm edge}\ne0$. Here $E_{m_*}$ abbreviates the energy at fixed $L$ and $\theta$ with offset $p$. Differentiating the crossing condition gives
\begin{equation}
\frac{\mathrm dg_*}{\mathrm dp}=-\frac{\partial_pE_{m_*}-\partial_p B_{\rm edge}}{\partial_gE_{m_*}-\partial_g B_{\rm edge}}.\label{eq:slope}
\end{equation}
For normalized physical and edge Bloch vectors $\psi$ and $\phi$, the Hellmann--Feynman identity \cite{Feynman1939} yields
\begin{equation}
\begin{aligned}
\partial_pE_{m_*}&=\sum_n|\psi_{n,3}|^2,&\partial_p B_{\rm edge}&=|\phi_3|^2,\\
\partial_gE_{m_*}&=\langle W\rangle_{m_*},&\partial_g B_{\rm edge}&=\sum_{r=2}^{8}v_r|\phi_r|^2.
\end{aligned}\label{eq:edge-weights}
\end{equation}
Here $\phi_r$ is the spacer component in one reference cell, and $\langle W\rangle_{m_*}$ is the expectation of $W$ in the active physical state.

For normalized physical and upper-edge states, denote the weights at spacer site $r=3$ by $q_{3,m}=\sum_n|\psi_{n,3}^{(m)}|^2$ and $b_3=|\phi_3|^2$. Figure~\ref{fig:control}(b) shows $R=b_3/\langle q_3\rangle_I$, with an equal arithmetic average over ranks. At every sampled offset, the edge weight exceeds each selected physical weight, so $\partial_p(B_5^+-E_m)=b_3-q_{3,m}>0$. This inequality explains the faster downward motion of the upper edge in Fig.~\ref{fig:control}(a) as $p$ decreases.

The spacer-weight difference can be traced to the interference between neighboring amplitudes and their energy detuning from spacer site $r=3$. The amplitudes from sites $r=2$ and $r=4$ add coherently at spacer site $r=3$. Its equation gives $\psi_{n,3}=t(\psi_{n,2}+\psi_{n,4})/(E-u_3)$ for $E\ne u_3$. Applying this relation to the physical and upper-edge states gives
\begin{equation}
q_{3,m}=\frac{t^2\sum_n|\psi_{n,2}^{(m)}+\psi_{n,4}^{(m)}|^2}{(E_m-u_3)^2},\quad
b_3=\frac{t^2|\phi_2+\phi_4|^2}{(B_5^+-u_3)^2}.\label{eq:coherent-weight}
\end{equation}
To quantify the interference, compare the coherent intensity $|\psi_2+\psi_4|^2$ with the incoherent sum $|\psi_2|^2+|\psi_4|^2$, taking the ratio after summing over cells and averaging each intensity over the selected ranks for the physical states. At $g/t=10.45$, the physical ratio changes from $0.08$ at $p=0$ to $0.32$ at $p=-4t$, while the corresponding edge ratio increases from $0.51$ to $1.96$. Between these two offsets, the edge interference changes from destructive to constructive, whereas the physical ratio of rank-averaged intensities remains below one. Lowering $p$ changes both the coherent numerators and the energy denominators in Eq.~\eqref{eq:coherent-weight}.

The resulting endpoint motion follows from Eq.~\eqref{eq:slope}, which also accounts for the relative response to $g$. Let $g_-$ and $g_+$ denote the entry and exit controls. At $p=0$, their slopes are $\mathrm dg_-/\mathrm dp\simeq-0.098$ and $\mathrm dg_+/\mathrm dp\simeq0.036$, giving $\partial_p(g_+-g_-)\simeq0.134$. Thus increasing $p$ moves the entry to smaller $g$ and the exit to larger $g$, widening the window near $p=0$. For finite offsets, the boundaries are recomputed with the active rank and edge determined at each crossing.

Scanning $g$ at each offset then determines how this relative motion changes the return-window endpoints [Fig.~\ref{fig:control}(c)]. The blue region marks complete returns ($G<0$). In this scan, the third-window width changes from $0.36t$ at $p=0$ to $0.53t$ at $p=+t$ and $0.16t$ at $p=-2t$. The third complete-return window closes at $p/t\simeq-3.61$. As $p$ decreases from $0$ to $-4t$, the first window widens slightly, the second narrows, and the third closes. At $p=-4t$, the first two complete returns have widths $1.04t$ and $0.44t$, compared with $0.91t$ and $0.74t$, respectively, at $p=0$. The original controls $g/t=0.6,6.5$ remain inside these windows, with all selected levels satisfying $G<0$ (Appendix~\ref{app:control-checks}).

At fixed $g/t=10.45$, Fig.~\ref{fig:control}(d) tests the associated change in spatial concentration through the size dependence of the typical IPR, averaged over all selected ranks and the same twelve phases at seven sizes from $L=801$ to $60885$. The IPR decreases with size at $p=0$ and $p=-2t$. At $p=-4t$, the typical IPR approaches approximately $0.48$ over the sampled sizes.

\subsection{Inverse endpoint design}\label{sec:inverse-design}
The single spacer offset in Fig.~\ref{fig:control} moves both endpoints of the third return along one local response direction. A second offset can provide an independent direction, allowing the window center and width to be adjusted separately. We seek prescribed entry and exit controls $(A,B)$ for the finite-chain rank window $I_L$ corresponding to the fixed IDS interval $I$, keeping the scan slopes $v_r$ fixed. For a set of spacer controls, write
\begin{equation}
u_r(g,\bm p)=w_r+v_rg+\sum_{a=1}^{d}C_{ra}p_a,\label{eq:design-offsets}
\end{equation}
where the dimensionless matrix $C$ specifies how the $d$ independent offsets act on the spacers. Let $\bm g(\bm p)=(g_-(\bm p),g_+(\bm p))^{\mathsf T}$ denote the complete-window endpoints. The endpoint response matrix $\mathsf D_{\sigma a}=\partial g_\sigma/\partial p_a$ measures how each offset shifts the entry and exit controls. At each endpoint $\sigma=-,+$ with active rank $m_\sigma$, define the physical and edge weights $q_{r,\sigma}=\sum_n|\psi_{n,r}^{(m_\sigma)}|^2$ and $b_{r,\sigma}=|\phi_{r,\sigma}|^2$. Under the same assumptions on the level and band-edge crossing, Eq.~\eqref{eq:slope} gives
\begin{equation}
\mathsf D_{\sigma a}=\frac{\partial g_\sigma}{\partial p_a}
=-\frac{\sum_{r=2}^{8}C_{ra}(q_{r,\sigma}-b_{r,\sigma})}
{\sum_{r=2}^{8}v_r(q_{r,\sigma}-b_{r,\sigma})}.
\label{eq:design-J}
\end{equation}
When $\mathsf D$ has row rank two, the offset responses span two independent directions in endpoint space. Provided all other selected levels remain strictly inside the reference band, the offsets then control the center $\bar g=(g_-+g_+)/2$ and width $\Omega=g_+-g_-$ independently to first order. The linear offset update with the smallest Euclidean norm is
\begin{equation}
\delta\bm p=\mathsf D^+\left[\begin{pmatrix}A\\B\end{pmatrix}-\bm g(\bm p)\right],
\label{eq:design-step}
\end{equation}
where $\mathsf D^+$ is the Moore--Penrose inverse. The endpoints are then recomputed, and the offsets are iteratively corrected toward the targets. The sampled interior controls are also checked for complete-window containment (Appendix~\ref{app:control-checks}).

\begin{figure}[!tbp]
\centering
\includegraphics[width=\columnwidth]{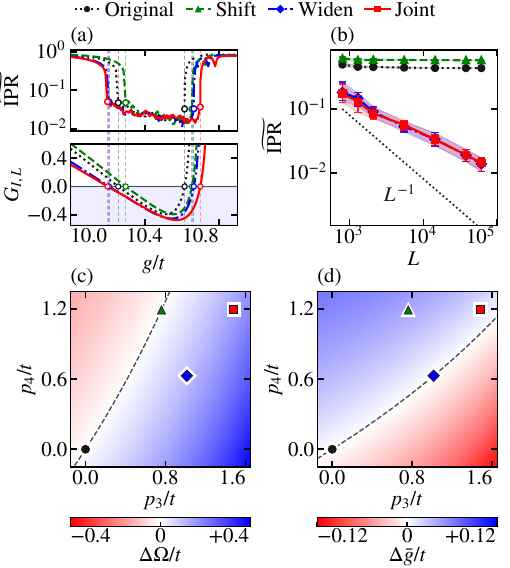}
\caption{(a) $\widetilde{\mathrm{IPR}}$ (upper logarithmic plot) and $G_{I,L}$ (lower linear plot) share a control axis at $L=14373$. Vertical guides mark $G=0$ boundaries in both plots. Positive $G$ is truncated above $0.6$. (b) $\widetilde{\mathrm{IPR}}$ at $g/t=10.23$ over all selected ranks and the same twelve phases as Fig.~\ref{fig:scan}(b), at seven sizes from $L=801$ to $60885$ with fixed offsets for each design. Shading and capped bars span the phase-specific geometric means. The dotted $L^{-1}$ guide has display value $0.1$ at $L=801$. (c),(d) Changes in width $\Delta\Omega/t$ and center $\Delta\bar g/t$ from Original at $L=14373$. Red and blue indicate negative and positive changes, respectively. Dashed contours mark zero change, and symbols identify the designs.}
\label{fig:inverse-design}
\end{figure}

For a concrete target, take $(A/t,B/t)=(10.20,10.70)$ and use independent offsets $p_3,p_4$ on spacer sites $r=3$ and $r=4$. This choice preserves both $\bm v$ and the bare crossing $u_2=u_6$ at $g/t=10.45$. The optimized offsets are approximately $p_3/t=1.48$ and $p_4/t=1.20$. The endpoint calculations use the unrounded optimized offsets.
Figure~\ref{fig:inverse-design}(a) compares Original with three designs. Joint extends the complete-return interval from approximately $(10.26t,10.62t)$ to $(10.20t,10.70t)$, increasing its width from $0.36t$ to $0.50t$. Shift targets a common endpoint displacement of $0.04t$ at fixed width, whereas Widen targets an outward displacement of $0.05t$ at each endpoint, increasing the width by $0.10t$ at fixed center. The IPR dips follow the displaced or widened $G<0$ intervals. All three designs retain the first two returns, narrowing the first and widening the second at the same size and phase (Appendix~\ref{app:control-checks}).

Figures~\ref{fig:inverse-design}(c) and~\ref{fig:inverse-design}(d) separate the width and center responses in the two-offset plane. Shift lies on the zero-width-change contour at positive center shift; Widen lies on the zero-center-change contour at increased width. The Joint design simultaneously increases the return-window width and shifts its center toward larger $g$.

At $g/t=10.23$, the selected finite-chain rank windows have $F_{\min}>0$ for Original and Shift, whereas they have $G<0$ for Widen and Joint, at every sampled size and phase (Appendix~\ref{app:control-checks}). With increasing system size at fixed offsets, the typical IPR varies weakly for Original and Shift but decreases for Widen and Joint [Fig.~\ref{fig:inverse-design}(b)], providing a finite-chain spatial signature of this sector change.

If the scan slopes can also be rescaled together, the entire return pattern can be shifted and stretched along the control axis by an exact affine transformation. Source endpoints $(g_-^{(0)},g_+^{(0)})$ are mapped to target values $(A,B)$ with $A<B$ by choosing
\begin{equation}
\begin{aligned}
s&=\frac{g_+^{(0)}-g_-^{(0)}}{B-A}>0, & c&=g_-^{(0)}-sA,\\
v'_r&=sv_r, & w'_r&=w_r+cv_r .
\end{aligned}\label{eq:design-affine}
\end{equation}
Then $H'(g)=H(sg+c)$ and $\mathcal N'_g(E)=\mathcal N_{sg+c}(E)$. The increasing map $g\mapsto(g-c)/s$ moves all returns together and transfers the original infinite-chain IDS certificates, with neighborhood radii divided by $s$. This exact certificate transfer follows from the affine operator identity; the independent-offset designs above establish numerical endpoint placement for finite-chain rank windows. Finite-chain source estimates yield finite-chain estimates of the mapped endpoints.

\section{Discussion and conclusions}\label{sec:discussion}
In the matched-dimer chain, an exact compression identity identifies the critical--localized boundaries with the edges of periodic reference bands. For the golden cell frequency, we prove on a common full-measure phase set for every finite $g$ that the physical spectrum is critical and purely singular continuous for $|\Delta(E,g)|<2$, and pure point with exponentially localized eigenfunctions for $|\Delta(E,g)|>2$. The absence of absolutely continuous spectrum rules out a conventional extended sector. Analytical upper and lower IDS bounds, combined with validated state counts, identify a common certified range of IDS labels and establish a C--L--C--L--C--L sequence for the same fixed IDS interval within six successive control neighborhoods. For finite chains, the unequal responses of the physical levels and reference-band edges to spacer offsets allow the return windows to be reshaped and their endpoints to be placed at prescribed controls.

Observing this alternation experimentally requires tracking the same spectral fraction as the control varies. Momentum-state lattices, superconducting circuits, and photonic mosaics offer ways to control local energies and resolve modes \cite{An2018,Xiao2021,LiCircuit2023,Gao2025}. In particular, programmable superconducting qubits have been used to realize a quasiperiodic mosaic lattice and identify critical states through their dynamics \cite{Huang2026}. At each static control $g$, frequency ordering identifies the modes in the fixed rank interval, and their normalized spatial intensities yield the typical IPR. Its variation across controls would reveal the alternating spatial concentration of this fraction. Spacer-energy shifts could test the control mechanism by comparing measured changes in physical levels with changes in the reference-band edges calculated from calibrated model parameters. Resolving the closely spaced levels near the third return favors smaller arrays (Appendix~\ref{app:finite-size}).

Beyond these static measurements, energy-filtered wave-packet spreading and long-time return probabilities could test whether successive critical--localized returns of the same selected spectral fraction are accompanied by corresponding changes in spreading and the persistence of local weight \cite{Ketzmerick1992,Ketzmerick1997,Last1996}.

\section*{Data availability}
The data, exact arithmetic certificates, and calculation and plotting code are available from the authors upon reasonable request.

\begin{acknowledgments}
This work is supported by the National Natural Science Foundation of China under Grants No. 11974205 and 61727801, the Key Research and Development Program of Guangdong province (2018B030325002), and the National Natural Science Foundation of China under Grant 62131002.
\end{acknowledgments}

\appendix
\section{Physical transfer matrix and periodic reference bands}\label{app:transfer}

At fixed physical energy, a cell-independent change of basis reduces the physical transfer matrix to the scalar recurrence in Eq.~\eqref{eq:sec}. Propagation through the matched dimer cancels the terms proportional to $J_n$ in its transfer matrix. Each spacer contributes an ordinary unit-hopping transfer matrix. In the boundary convention of Sec.~\ref{sec:transfer}, this gives
\begin{align}
M_n&=P(D_\infty+B_0/J_n),&
P&=S_{\kappa-1}\cdots S_2,\nonumber\\
D_\infty&=\begin{pmatrix}-2E&1\\-1&0\end{pmatrix},&
S_r&=\begin{pmatrix}E-u_r(g)&-1\\1&0\end{pmatrix},\nonumber\\
B_0&=(E,1)^{\mathsf T}(E,-1).
\end{align}
Thus $A=PD_\infty$ satisfies $\det A=1$, and $B=PB_0$ has rank one. With $z=(1,E)^{\mathsf T}$, the identity $B=-(Az)(E,-1)$ gives $Bz=0$, $BAz=\tau Az$, and $\det Q=-\tau$ for $Q=(Az,z)$. If $\tau=0$, then $Az=\lambda z$ for a nonzero real $\lambda$, so $\Delta=\lambda+\lambda^{-1}$ and $|\Delta|\ge2$. For $\tau\ne0$, Cayley--Hamilton gives
\begin{equation}
Q^{-1}M_nQ=
\begin{pmatrix}\Delta+\tau/J_n&1\\-1&0\end{pmatrix}.
\end{equation}
The coordinates $Q^{-1}X_n=(y_n,-y_{n-1})^{\mathsf T}$ therefore obey Eq.~\eqref{eq:sec} at each fixed physical energy.

The same discriminant $\Delta$ determines the periodic reference bands through the Bloch relation $\Delta=2\cos K$. For the periodic compression, let $T_{\rm sp}$ be the open spacer chain, $C=\det(E-T_{\rm sp})$, and $C_{\rm L},C_{\rm R}$ the characteristic polynomials obtained by deleting its last and first sites, respectively. The spacer recurrence gives $(P_{11},P_{12},P_{21})=(C,-C_{\rm R},C_{\rm L})$, hence $\Delta=-2EC+C_{\rm L}+C_{\rm R}$. In a Bloch convention with phase $e^{iK}$, the compressed Hamiltonian is
\begin{equation}
h_{\rm b}=\begin{pmatrix}0&\ell^\dagger\\\ell&T_{\rm sp}\end{pmatrix},
\qquad \ell=\frac{-e_1+e^{iK}e_{\kappa-2}}{\sqrt 2},
\end{equation}
where $e_j$ are spacer basis vectors. In the bordered determinant $EC-\ell^\dagger\operatorname{adj}(E-T_{\rm sp})\ell$, the diagonal endpoint cofactors are $C_{\rm L},C_{\rm R}$ and the endpoint-to-endpoint cofactor is one. Thus $\det(E-h_{\rm b})=EC-\tfrac12(C_{\rm L}+C_{\rm R})+\cos K$, yielding Eq.~\eqref{eq:backbone}. This polynomial identity includes spacer eigenenergies.

\section{Spectral classification and Lyapunov exponent}\label{app:spectral}

Appendix~\ref{app:proof} excludes absolutely continuous spectrum, ruling out a conventional extended phase. It also excludes point spectrum in the open reference bands, establishing pure singular continuity under the specified phase conditions. Appendix~\ref{app:lyapunov} derives the exact Lyapunov exponent on the physical spectrum, which equals zero inside the open reference bands. Together with pure singular continuity, this identifies the critical phase. Appendix~\ref{app:localization} uses the positive exterior exponent and Green-function estimates to establish pure-point spectrum with exponentially decaying eigenfunctions.

\subsection{Singular-continuous spectrum inside the reference bands}\label{app:proof}

For irrational $\alpha$, choose weak bonds in both tails with $\sum_j|J_{n_j}|<\infty$. Cutting only these hoppings changes $H$ by a trace-class operator of trace norm at most $2\sum_j|J_{n_j}|$ and leaves a direct sum of finite chains. Kato--Rosenblum invariance then excludes absolutely continuous spectrum for every phase and finite $g$ \cite{Kato1957,Rosenblum1957,SimonSpencer1989}.

Set $\omega=2\pi\alpha$. To exclude normalizable eigenstates, use the following phase set, which excludes only countably many phases:
\begin{equation}
\begin{split}
\Theta_*=\{\theta:\ &\cos(\theta+n\omega)\ne0,\\
 &2\theta+m\omega\notin2\pi\mathbb Z\quad (n,m\in\mathbb Z)\}.
\end{split}
\end{equation}
If $|\Delta|<2$, a physical $\ell^2$ eigenvector gives an $\ell^2$ solution $y$ of Eq.~\eqref{eq:sec} by the invertible change of basis within each cell. This solution is nonzero since vanishing boundary data would force the entire eigenvector to vanish by the site recurrences. Write $a=\Delta$, $b=\tau$, $d(x)=2\cos x-a$, and let $f(x)=\sum_n y_ne^{inx}$ in $L^2$. Multiplying the scalar equation by $J_n$ before Fourier transformation gives
\begin{equation}
e^{i\theta}d(x+\omega)f(x+\omega)
+e^{-i\theta}d(x-\omega)f(x-\omega)=bf(x).
\end{equation}
Following the Fourier argument of Ref.~\cite{WangZhou2025}, set $u(x)=(f(x),e^{-i\theta}f(x-\omega))^{\mathsf T}$ and $F=(u,\bar u)$. The Fourier equation gives
\begin{equation}
\begin{split}
D(x)F(x)&=F(x+\omega)\operatorname{diag}(e^{i\theta},e^{-i\theta}),\\
D(x)&=\begin{pmatrix}
b/d(x+\omega)&-d(x-\omega)/d(x+\omega)\\
1&0
\end{pmatrix}.
\end{split}
\end{equation}
The determinant equation makes $|\det F(x)d(x-\omega)d(x)|$ invariant under rotation and thus constant almost everywhere. Since $\det F\in L^1$, the constant must vanish: a nonzero constant would give a nonintegrable inverse-distance singularity at a simple zero of $d$. The zero set of $u$ is invariant up to null sets because $D$ is invertible almost everywhere, so $f\ne0$ implies $u\ne0$ almost everywhere. Consequently $u=h\bar u$ for a measurable $|h|=1$, and $h(x+\omega)=e^{-2i\theta}h(x)$. This contradicts the rotation eigenvalues $e^{im\omega}$ on $L^2$ for $\theta\in\Theta_*$. The phase exclusion is independent of $E$ and $g$, so the absence of point and absolutely continuous spectrum gives purely singular-continuous physical spectrum in the open reference bands on $\Theta_*$.

Reference-band edge energies also carry no point spectrum on $\Theta_*$. At $|\Delta|=2$ with $\tau\ne0$, the same Fourier argument applies. The zeros of $d$ are quadratic, so the inverse-distance singularity in the determinant equation is then of order two. At $\tau=0$, the transfer matrices are upper triangular in a basis beginning with $z$ from Appendix~\ref{app:transfer}, with diagonal entries both equal to $1$ or both equal to $-1$. An $\ell^2$ solution must have zero lower coordinate and then zero upper coordinate. Thus no normalizable eigenstates occur at the reference-band edge energies for $\theta\in\Theta_*$. The spectral classification in Sec.~\ref{sec:spectral} concerns the two open sectors.

\subsection{Lyapunov exponent on the spectrum}\label{app:lyapunov}

Pure singular continuity excludes an absolutely continuous extended sector and pure-point localization, but does not by itself imply a zero Lyapunov exponent. On the physical spectrum, the exact exponent derived below is zero in the open reference bands, completing their characterization as critical, and is positive strictly outside them, where it enters the localization proof.

Fix a real energy $E$ and a finite control $g$. For the physical Jacobi equation with on-site energies $b_j$ and hoppings $c_j$ from $j$ to $j+1$, use the analytic numerator $U_j=\begin{pmatrix}E-b_j&-c_{j-1}\\c_j&0\end{pmatrix}$. Within a cell $c_0=b_0=b_1=J(x)=2\cos x$, and the regular hoppings are one. The cell numerator $C_E=U_{\kappa-1}\cdots U_0$ then satisfies
\begin{equation}
C_E(x)=J(x)A+B,
\qquad \det C_E(x)=J(x)^2.\label{eq:cell-numerator}
\end{equation}
This product also holds at the hopping zeros. Repeated multiplication of $C_E(x)$ as the phase advances by $2\pi\alpha$ defines the cell cocycle. Let $f(\varepsilon)=L(\alpha,C_E(\cdot+i\varepsilon))$ be its phase-averaged largest Lyapunov exponent per cell and $h=\log\rho(A)\ge0$. As $\varepsilon\to+\infty$, the rescaled numerator $e^{-\varepsilon}C_E(x+i\varepsilon)$ converges to $e^{-ix}A$ in an analytic strip norm. Continuity of the complex analytic exponent at irrational frequency gives $f(\varepsilon)-\varepsilon\to h$ \cite[Theorem~1.5]{AvilaJitomirskayaSadel2014}. The zeros of $J(z)=2\cos z$ lie on the real axis, so $C_E(z)/J(z)$ is analytic in a neighborhood of each positively shifted circle. Its determinant is one by Eq.~\eqref{eq:cell-numerator}. For this determinant-one cocycle, the right derivative of its Lyapunov exponent with respect to the imaginary phase shift, called the acceleration, is integer in our radian convention \cite{Avila2015,AvilaJitomirskayaSadel2014}. Jensen's identity gives $\int\log|J(x+i\varepsilon)|\,\mathrm dx/(2\pi)=|\varepsilon|$, so $f$ also has integer slopes for $\varepsilon>0$. Evenness, convexity and the large-shift slope one restrict these slopes to $0$ or $1$. The asymptote then gives
\begin{equation}
f(\varepsilon)=\max\{f(0),|\varepsilon|+h\},
\qquad f(0)\ge h .
\end{equation}

In the acceleration argument, the normalized cocycle has exponent $f(\varepsilon)-\varepsilon$ for $\varepsilon>0$; continuity and real-axis regularity are applied directly to the analytic numerator $C_E$, which may be singular. If $f(0)>h$, then $f$ is constant near zero. The sum of the two exponents on the real axis is the mean log determinant $2\int\log|J(x)|\,\mathrm dx/(2\pi)=0$. The exponents $f(0)$ and $-f(0)$ therefore have a strict gap. The regularity theorem for analytic complex matrix cocycles gives a dominated cell splitting $E^u(x)\oplus E^s(x)$ \cite[Theorem~1.2]{AvilaJitomirskayaSadel2014}: propagation along $E^s$ becomes exponentially smaller than along $E^u$, uniformly in phase. Each cell matrix maps the unstable line onto the next unstable line and maps the stable line into the next stable line.

At a cut $i$ inside a cell, let $P_i(x)$ be the product of the site factors before the cut and $R_i(x)$ the product after it, so that $C_E=R_iP_i$. Define
\begin{equation}
\begin{aligned}
E_i^u(x)&=P_i(x)E^u(x),\\
E_i^s(x)&=\{v:R_i(x)v\in E^s(x+\omega)\}.
\end{aligned}
\end{equation}
The map $R_i$ sends $P_iE^u$ nontrivially onto $E^u(x+\omega)$. Projecting $R_i$ onto this line along $E^s(x+\omega)$ gives a continuous rank-one map with kernel $E_i^s$. Thus the two lines vary continuously and remain separated by an angle bounded away from zero. This construction uses no inverse and remains valid at $J=0$. The partial-cell product identities preserve the splitting under each site factor, which acts nontrivially on the unstable line because the full cell product does so.

A product $T_k$ of $k$ site factors contains partial-cell products at the two ends, separated by $n$ complete-cell products. Compactness and the finitely many cuts give uniform upper bounds for these boundary factors and positive lower bounds on their unstable lines. Cell domination therefore yields, for unit vectors $v_s,v_u$ in the stable and unstable lines at the starting cut,
\begin{equation}
\frac{\|T_kv_s\|}{\|T_kv_u\|}\le C e^{-cn},
\qquad n=k/\kappa+O(1),
\label{eq:site-domination}
\end{equation}
with $C,c>0$ independent of the starting phase, cut and length. The unstable restriction also has a positive uniform lower bound at each fixed length. This proves single-site domination for the continuous Jacobi numerator cocycle $B^E=U_j$ on $\mathbb T\times\mathbb Z_\kappa$. The base advances one site at a time and rotates by $\omega$ after each completed cell, so all orbits are dense. The continuous invariant lines, uniform domination and nonvanishing unstable restriction established above satisfy the splitting conditions of the singular-Jacobi criterion, including at hopping zeros. The singular-Jacobi spectral criterion then places $E$ outside the physical spectrum \cite[Theorem~7]{AlkornZhang2022}.

For $E\in\Sigma_g$, the spectral criterion excludes $f(0)>h$; together with the lower bound above, this gives $f(0)=h$. Along the common rotation $x\mapsto x+\omega$, the identity $C_E=JM$ makes each $N$-cell numerator product equal to the physical transfer product multiplied by $\prod_{j=0}^{N-1}J(x+j\omega)$. Phases whose orbits meet a zero of $J$ form a null set, and $\log|J|$ is integrable. Homogeneity of the matrix norm therefore gives the exact normalization
\begin{equation}
L(\alpha,M)=f(0)-\int_0^{2\pi}\log|J(x)|\,\frac{\mathrm dx}{2\pi}=h.
\label{eq:physical-LE-normalization}
\end{equation}
This is the physical exponent per cell. Dividing by $\kappa$ and using $\det A=1$ and $\Tr A=\Delta$ gives Eq.~\eqref{eq:exactLE}.

\subsection{Exponential localization outside the reference bands}\label{app:localization}

The positive Lyapunov exponent obtained in Appendix~\ref{app:lyapunov} does not by itself establish localization. A fixed-energy reduction to the matched-dimer parent model and Green-function estimates give exponential decay of physical eigenfunctions, completing the localized part of the C--L classification.

For the golden cell frequency $\alpha=(\sqrt5-1)/2$, $\beta=\alpha/2$ satisfies $\|k\beta\|_{\mathbb T}\ge c(1+|k|)^{-\sigma}$ for $k\ne0$ with $c>0$ and $\sigma>1$. Let $\Theta_0(\beta)$ consist of phases with $\cos(\theta+2\pi n\beta)\ne0$ for every integer $n$. With $\varphi=\theta/(2\pi)$, define
\begin{equation}
\begin{split}
\Theta_{\rm loc}={}&\{\theta\in\Theta_*\cap\Theta_0(\beta):\exists\eta>0,\\
&\|2\varphi-k\beta\|_{\mathbb T}
\ge\eta(1+|k|)^{-\sigma}\ \ \forall k\in\mathbb Z\}.
\end{split}\label{eq:thetaloc}
\end{equation}
Here $\|\cdot\|_{\mathbb T}$ is distance to the nearest integer. Borel--Cantelli and exclusion of the countable resonances and cosine zeros give full measure. The cosine condition covers the full parent site-frequency orbit. The set is invariant under $\theta\mapsto\theta+\pi$ and common to all energies and controls.

Consider a polynomially bounded generalized eigenfunction at $|a|=|\Delta|>2$. For $b=\tau\ne0$, map its scalar equation at fixed energy to the matched-dimer parent of Ref.~\cite{WangZhou2025} with
\begin{equation}
\lambda=\frac{2|b|}{a^2-4},\qquad
e=-\frac{a\lambda}{2},\qquad |e|>\lambda .
\end{equation}
The parent has intercell hopping $\lambda$ and equal dimer on-site energies and internal hopping $\cos(\theta_p+n\omega)$, where $\theta_p=\theta$ for $b>0$ and $\theta_p=\theta+\pi$ for $b<0$. Its phase-independent matrix is $A_p=\begin{pmatrix}-2e/\lambda&1\\-1&0\end{pmatrix}$. The change of basis from Appendix~\ref{app:transfer}, with $z_p=(\lambda,e)^{\mathsf T}$ and $Q_p=(A_pz_p,z_p)$, is invertible since $\det Q_p=\lambda^2-e^2$. It gives the scalar trace $-2e/\lambda+(e^2-\lambda^2)/[\lambda\cos(\theta_p+n\omega)]$. Since $2(e^2-\lambda^2)/\lambda=|b|$, the parent scalar recurrence is Eq.~\eqref{eq:sec}. The changes of basis are independent of cell position at fixed energy and preserve polynomial bounds. A nonzero generalized solution thus places $e$ in the parent spectrum \cite{Teschl2000}.

Write $r=\rho(A_p)=(|e|+\sqrt{e^2-\lambda^2})/\lambda$, $L_p=\tfrac12\log r$, $\mu=\tfrac12\log(\lambda/2)$, and $\ell=L_p+\mu$. Appendix~\ref{app:lyapunov} applied to the rescaled parent gives $L_p$ per parent site on its spectrum: the numerator intercept $\log(\lambda/2)+\log r$ loses $\log(\lambda/2)$ upon division by $\lambda\cos x$, leaving $\log r$ per dimer. The following estimates are at fixed $E,g$ and $\theta\in\Theta_{\rm loc}$; decay constants may also depend on the generalized eigenfunction.

For a complete parent dimer block $[s,t]$ of length $k=t-s+1=2m$, write $P_k=\det(e-H_{p,[s,t]})$ and $\mathcal G_{[s,t]}=(e-H_{p,[s,t]})^{-1}$. With $c_j$ denoting the parent hoppings, Cramer's rule gives the numerator of $|\mathcal G_{[s,t]}(y,s)|$ as $|\det(e-H_{p,[y+1,t]})|\prod_{j=s}^{y-1}|c_j|$: the determinant and hopping product have lengths $t-y$ and $y-s$, respectively. For the right endpoint, the corresponding factors use $[s,y-1]$ and $\prod_{j=y}^{t-1}|c_j|$, with lengths $y-s$ and $t-y$. Thus, at either endpoint $\xi\in\{s,t\}$ and distance $d=|y-\xi|$, the two lengths are $k-1-d$ and $d$. Uniform subadditive bounds for the numerator matrices and unique ergodicity applied to truncated log hoppings bound these factors by $C_\delta e^{(k-1-d)(\ell+\delta)}$ and $C_\delta e^{d(\mu+\delta)}$. A common $C_\delta\ge1$ can be chosen for both starting parities and all lengths, depending on the fixed $e,\lambda,\alpha$ and $\delta$, but not on phase or block position.

Fix $0<\varepsilon<L_p$ and $0<\delta<\varepsilon/56$. Whenever $|P_k|>e^{(k+1)(\ell-\varepsilon/8)}$, multiplying the two numerator bounds and dividing by this lower bound, with $\ell-\mu=L_p$, gives
\begin{equation*}
\begin{aligned}
\log|\mathcal G_{[s,t]}(y,\xi)|
&\le -L_p d+\delta(k-1)\\
&\quad +(k+1)\varepsilon/8+C_0,
\end{aligned}
\end{equation*}
where $C_0=2\log C_\delta-2\ell$ is independent of $k,d$, phase and block position. At $d\ge k/7$, the choice $\delta<\varepsilon/56$ leaves a positive margin to absorb this constant, giving decay at rate $L_p-\varepsilon$ for sufficiently large $k$.

The required large determinants follow from a lower bound at the same block length. In the symmetric and dark dimer basis, each internal hopping enters only one diagonal entry. The coefficient of their product is $(-2)^m D_m(e)$, where $D_m$ is the open dark-chain determinant with hopping $-\lambda/2$. Since each cosine contributes a Fourier coefficient of modulus $1/2$, $|\widehat P_{2m}(m)|=|D_m(e)|$. The dark-chain recurrence gives $|D_m(e)|=(\lambda/2)^m\sum_{j=0}^{m}r^{m-2j}$, hence
\begin{equation}
\|P_{2m}\|_\infty\ge|\widehat P_{2m}(m)|\ge e^{2m\ell}.
\end{equation}
Reflection gives a degree-$m$ polynomial in $\cos[2\pi(\varphi+(m-1)\beta)]$ \cite[Lemma~3.14]{WangZhou2025}. For $m+1$ cosine nodes, $\varepsilon'$-uniformity means that every Lagrange interpolation weight has magnitude at most $e^{m\varepsilon'}$ throughout $[-1,1]$. If the polynomial values at these nodes are all at most $e^{(2m+1)(\ell-\varepsilon/8)}$ in modulus, Lagrange interpolation gives an upper bound $(m+1)e^{(2m+1)(\ell-\varepsilon/8)+m\varepsilon'}$ on its maximum magnitude over $[-1,1]$. For $0<\varepsilon'<\varepsilon/8$ this contradicts the lower bound at large $m$. This supplies the same-scale step in place of Lemma~3.19 of Ref.~\cite{WangZhou2025}.

Translate a nonzero dimer of the parent generalized eigenfunction to the origin: $u'(j)=u(j+2n_0)$ has phase $\theta'=\theta_p+n_0\omega$ and $u'(y_0)\ne0$ for some $y_0\in\{0,1\}$. This fixed shift preserves $\Theta_{\rm loc}$, with an adjusted Diophantine constant depending on $n_0$ but not on the block length or observation distance.

For large $y>0$ let $q$ be the largest continued-fraction denominator of $2\beta=\alpha$ with $q\le y/8$. Successive denominators of the golden frequency have ratio less than two at these scales. Thus $8q\le y<16q$. Set $m=6q$ and $k=2m$. Take the integer node groups
\begin{equation}
[0,2q]\quad\text{and}\quad
[\lfloor y/2\rfloor+1-2q,\lfloor y/2\rfloor+2q].
\end{equation}
The $m+1$ nodes $\varphi_j=\theta'/(2\pi)+(2j-1)\beta$ are the center phases of blocks $[2j-m,2j+m-1]$. The first group contains $y_0$ in each block and the second contains $y$. In both groups the distances from the respective site to the two endpoints are at least $2q-2$, hence at least $k/7$ for large $q$.

The nodes form six $q$-blocks and one extra node, with phase increment $2\beta=\alpha$. Factoring cosine differences into phase-sum and phase-difference sine factors allows the logarithmic sine-sum estimate of Ref.~\cite[Lemma~2.1]{WangZhou2025} to be applied on each block. After omitting the smallest sine, its sum is $-(q-1)\log2+O(\log q)$, uniformly in the starting phase. The frequency and phase conditions bound the omitted denominator terms and the extra node by logarithmic errors, since $|i-j|,|i+j-1|=O(q)$; the self difference is excluded and the phase condition includes $i+j-1=0$. Thus the logarithm of each Lagrange basis product is at most $O(\log q)=o(m)$ uniformly on $[-1,1]$, giving $\varepsilon'$-uniformity for any fixed $\varepsilon'>0$ at sufficiently large $q$.

If a block in the first group has determinant larger than the threshold in the Green-function estimate, both endpoint Green functions decay at rate $L_p-\varepsilon>0$ away from $y_0$. The Green representation, with exterior hoppings $\lambda$ and polynomially bounded exterior amplitudes, would then force the fixed nonzero amplitude $u'(y_0)$ to tend to zero as $y$ grows. Thus all first-group determinants are small at large $y$. The interpolation obstruction with $\varepsilon'<\varepsilon/8$ forces at least one second-group determinant above the same threshold. Its Green representation gives exponential decay of $u'(y)$ since the endpoints are at distances of order $y$. Reflection $j\mapsto1-j$ sends $\theta'$ to $-\theta'$, which preserves the arithmetic conditions, and yields the other tail.

The changes of basis, which are independent of cell position, transfer this decay to $X_n$ of the physical chain. The components $(\psi_{n,2},\psi_{n,1})^{\mathsf T}=P^{-1}X_{n+1}$ and the remaining spacers follow by unit-hopping recurrences, with no inverse weak bond. At the exceptional energies $\tau=0$, $Az=\rho z$ with $|\rho|\ne1$. In a basis beginning with $z$, $A$ is upper triangular with diagonal $(\rho,\rho^{-1})$ and $B$ is strictly upper triangular. Two-sided polynomial boundedness forces the lower coordinate to vanish first and then the upper one. Thus no nonzero polynomially bounded solution occurs at these energies. The generalized-eigenfunction expansion for the physical Jacobi operator gives a pure-point restriction to $|\Delta|>2$ with exponentially decaying eigenfunctions.

\section{Certification of the fixed IDS interval}\label{app:IDS}

The IDS is continuous, with no jump at any energy. For almost every phase, all hoppings are nonzero and the Jacobi recurrence bounds the dimension of any eigenspace by two. The phase average of the trace of its spectral projection on $L$ sites equals $L$ times the size of the IDS jump at that energy, and is bounded by two. Dividing by $L$ and taking $L\to\infty$ excludes any such jump.

To obtain reliable IDS bounds, the finite-chain counts must account for numerical errors. Let $a_i,b_i$ be the energy endpoints, specified as exact decimal values, used in the counts at $g_i$. At each sampled phase $\phi_j$, we count the negative eigenvalues of the matrix reconstructed from the tridiagonal factorization of $H_L(g_i,\phi_j)-\widehat E\,\mathrm{Id}$. Input enclosures and rounding-error estimates bound the backward errors by $\eta_{j,a}$ and $\eta_{j,b}$ at the comparison energies $\widehat E_a$ and $\widehat E_b$. The one-sided energy margins at $g_i$ satisfy
\begin{align}
\widehat E_a-(a_i+\epsilon_\uparrow)&>\eta_{j,a},\nonumber\\
(b_i-\epsilon_\uparrow)-\widehat E_b&>\eta_{j,b}.
\end{align}
Weyl's inequality converts these negative-eigenvalue counts into an upper count at $a_i+\epsilon_\uparrow$ and a lower count at $b_i-\epsilon_\uparrow$. Unresolved counts contribute $L$ to the upper sum or zero to the lower sum.

These bounds at the control centers extend to finite neighborhoods by allowing for the corresponding energy shifts. For $|g-g_i|\le d=5\times10^{-6}t$, the perturbation norm is at most $d$ since $\|W\|=1$. Applying Weyl's inequality to the ordered finite-chain energies and taking the ergodic limit gives
\begin{equation}
\mathcal N_g(a_i-d)\le\mathcal N_{g_i}(a_i),
\qquad
\mathcal N_g(b_i+d)\ge\mathcal N_{g_i}(b_i).
\end{equation}
The outward-rounded enclosures in Table~\ref{tab:certificates} contain $[a_i-d,b_i+d]$, so the IDS bounds apply at the displayed endpoints throughout each control neighborhood. Over each energy--control rectangle, the discriminant polynomial lies between the smallest and largest coefficients in its tensor-product Bernstein representation. This property gives the bounds in Table~\ref{tab:occupation} throughout the rectangle. The limiting IDS bounds are $\max U_a=0.557661$ at $L_2$ and $\min L_b=0.575282$ at $C_1$; inward rounding gives the common range $\mathcal J$ in Sec.~\ref{sec:certification}.

\begin{table}[H]
\caption{Energy enclosures and IDS bounds underlying Table~\ref{tab:occupation}. The enclosures are rounded outward, with $\mathcal N_g(a)\le U_a$ and $\mathcal N_g(b)\ge L_b$ throughout each control neighborhood.}
\label{tab:certificates}
\setlength{\tabcolsep}{4pt}
\begin{ruledtabular}
\begin{tabular}{cccc}
$g_i/t$ & $[a,b]/t$ & $U_a$ & $L_b$ \\
\hline
$0.6$ & $[-0.153331,-0.143533]$ & $0.536714$ & $0.575282$ \\
$3.5$ & $[-0.620005,-0.599995]$ & $0.556242$ & $0.579324$ \\
$6.5$ & $[-0.208844,-0.206367]$ & $0.514328$ & $0.592309$ \\
$8.5$ & $[1.67502,1.67511]$ & $0.557661$ & $0.591545$ \\
$10.45$ & $[3.573186,3.573415]$ & $0.556804$ & $0.580432$ \\
$12$ & $[3.82999,3.85001]$ & $0.556886$ & $0.584691$ \\
\end{tabular}
\end{ruledtabular}
\end{table}

\section{Parameter selection and numerical checks}\label{app:scaling}

\subsection{Parameter selection and five-site reentry}\label{app:selection}

\begin{figure}[!htbp]
\includegraphics[width=\columnwidth]{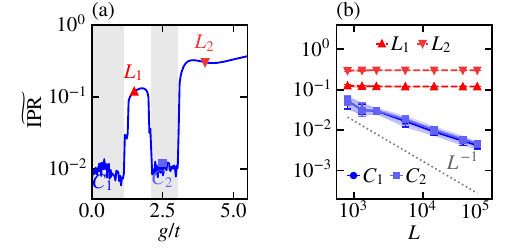}
\caption{Five-site reentry in the fixed IDS interval $I_5=[0.4225,0.4355]$. (a) $\widetilde{\mathrm{IPR}}$ at $L=14375$ and $\theta=0.37$; gray regions mark complete band containment. (b) Size dependence at the four marked controls, using the phase averaging of Fig.~\ref{fig:scan}(b) and the nearest chain lengths compatible with complete five-site cells. Shading and capped bars show the range across phases.}
\label{fig:five-common}
\end{figure}

The parameters used in the main text were identified by an exploratory search for repeated returns in small open chains. The selected spacer offsets were rounded to two decimal places, and the final IDS interval $I=[0.558,0.571]$ was fixed for the larger-chain calculations. Both the unrounded and rounded parameter sets retain three complete returns in the finite-chain comparison. The IDS certificates apply to the rounded model.

To examine whether reentry persists at a different cell length, we consider a five-site chain with $\bm v=(0,0,1,1,-1)$, $\bm w/t=(0,0,-1.75,2,3.75)$, and a fixed IDS interval $I_5=[0.4225,0.4355]$ of weight $0.013$, matching the nine-site example. Finite-chain calculations at $g/t=0.4,1.5,2.5,4$ show that every selected rank satisfies the corresponding C--L--C--L sector condition over all sampled sizes and phases. The typical IPR decreases overall with size at the two critical controls and varies weakly at the localized controls [Fig.~\ref{fig:five-common}].

\subsection{Window weight and robustness of the finite-chain results}\label{app:finite-size}
Increasing the selected spectral weight makes complete alternation more restrictive, because every additional level must satisfy the sector condition at all six controls. To examine how much the window can be widened, we vary its lower filling $f_-$ for each tested weight $\mathcal W$, seeking an interval $[f_-,f_-+\mathcal W]$ that meets all six conditions. The scan uses $L=2097,5490,14373$ and four phases, $\theta=0.17,0.37,1.31,2.73$. On the sampled filling grid, a window of weight $0.02$ satisfies these conditions in all twelve size--phase combinations, whereas no sampled window of weight $0.03$ does. These finite-chain results show that the chosen weight $\mathcal W=0.013$ is not the only one that supports complete alternation. The $0.02$ result concerns sampled finite chains; the infinite-chain IDS bounds in Sec.~\ref{sec:certification} certify weights approaching $0.0175$ from below, without setting an upper limit on possible alternating weights.

For the fixed interval $I$ used in the main text, restricting the IPR fits to larger chains tests whether the decrease at the critical controls persists beyond the smaller sizes. Using the same unweighted fit in Eq.~\eqref{eq:D2}, the effective exponents $D_2^{\rm fit}$ for $C_1,C_2,C_3$ are $(0.585, 0.566, 0.567)$ over all seven sizes ($L=801$ to $60885$) and $(0.603, 0.588, 0.585)$ over the largest three ($L=14373$ to $60885$). The two-point slopes between $L=37629$ and $60885$ are $(0.649, 0.633, 0.610)$. The IPR therefore continues to decrease over the larger sizes, although the slope increases as the smaller sizes are excluded. This dependence on the fitted range supports interpreting $D_2^{\rm fit}$ as a finite-size effective exponent.

The dependence on the fitted size range should be distinguished from the effect of phase sampling. The comparison uses uniformly spaced phases $\theta_k=0.43+2\pi k/K$, $k=0,\ldots,K-1$, with equal weight assigned to all selected ranks and phases. Increasing $K$ from $12$ to $24$ changes the typical IPR by at most $3.8\%$ across the tested critical configurations and sizes. The corresponding changes in $D_2^{\rm fit}$ are at most $0.004$ for the seven-size fits and $0.011$ for the largest-three-size fits. The two-point slopes between the largest sizes change more noticeably, indicating greater sensitivity to phase sampling than the fits over broader size ranges.

Numerical reliability is checked at all three critical controls $C_1,C_2,C_3$ by varying the eigenvalue tolerance at $L=60885$ and $\theta=0.17,1.31,2.73$. The closely spaced levels at $C_3$ provide a sensitive test, yet changing the solver tolerance has little effect on the window-averaged IPR or its fitted size dependence. Independent full-window solves at $\theta=0.43$ cover all configurations in Figs.~\ref{fig:scan}(b), \ref{fig:control}(d), and \ref{fig:inverse-design}(b) at $L=14373$, with additional checks of $C_3$, $p=-2t$, and Joint at $L=60885$. At the same size, parameters, and phase, the relative difference in the typical IPR between the independent and original calculations is less than $0.05\%$.

\subsection{Spacer control and inverse-design checks}\label{app:control-checks}

\begin{figure}[!t]
\includegraphics[width=\columnwidth]{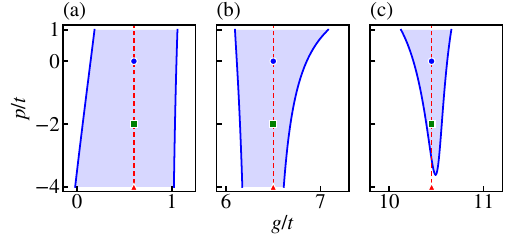}
\caption{Complete-return regions under the spacer offset $w_3\mapsto w_3+p$ for the first (a), second (b), and third (c) returns. Blue shading denotes $G_{I,L}<0$; white includes partial or no returns. Red vertical dashed lines mark the unperturbed critical controls $g/t=0.6,6.5,10.45$, respectively. Circles, squares, and triangles mark $p=0,-2t,-4t$, as in Fig.~\ref{fig:control}(c). All panels use the same horizontal span, $I=[0.558,0.571]$, and an open chain with $L=14373$ and $\theta=0.37$.}
\label{fig:all-return-control}
\end{figure}

\begin{figure}[!t]
\includegraphics[width=\columnwidth]{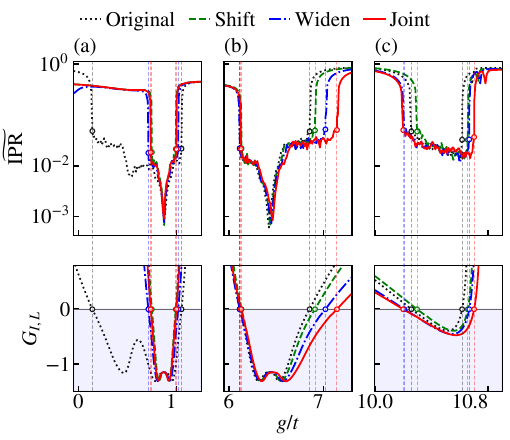}
\caption{The first (a), second (b), and third (c) returns for the four configurations in Fig.~\ref{fig:inverse-design}(a), using $I=[0.558,0.571]$, $L=14373$, and $\theta=0.37$. Upper plots show $\widetilde{\mathrm{IPR}}$, geometrically averaged over the selected ranks, on a logarithmic axis; lower plots show $G_{I,L}$, with shading at $G_{I,L}<0$. Vertical dashed guides and open circles mark the complete-return endpoints. Values of $G_{I,L}>0.8$ are outside the displayed range.}
\label{fig:all-return-design}
\end{figure}

For the unperturbed chain, the three return windows remain at nearly the same control values across the tested sizes up to $L=60885$ and phases $\theta=0.17,1.31,2.73$. Independent eigenvalue checks confirm the boundary crossings. At the sampled interior controls, all selected levels lie inside the reference bands ($G<0$), while at $L_1,L_2,L_3$ they all lie in the exterior.

Among the intervening spacers, site $r=3$ gives the largest absolute third-window width response in the boundary-derivative comparison at $L=14373$. For this perturbation [Fig.~\ref{fig:control}], the offset at which the complete third return disappears remains nearly unchanged across the tested sizes $L=2097,5490,14373$ and phases $\theta=0.17,1.31,2.73$.

The three return windows respond differently to the same spacer offset [Fig.~\ref{fig:all-return-control}]. As $p$ decreases from $0$ to $-4t$, the first window widens slightly, the second narrows, and the third closes. The original controls $g/t=0.6,6.5$ remain inside the first two complete-return regions at every sampled offset in this range.

The tested designs satisfy the response and crossing conditions in Sec.~\ref{sec:inverse-design}. With the unrounded offsets, all selected levels lie inside the reference bands at the sampled interior controls and outside them at the sampled exterior controls. With these offsets held fixed, the endpoints of Joint, Shift, and Widen remain nearly unchanged across the tested sizes and phases. Independent eigenvalue calculations give consistent endpoint estimates, and direct solutions at the sampled off-grid offsets agree with the interpolated results in Fig.~\ref{fig:inverse-design}(c,d).

At the fixed control $g/t=10.23$, the endpoint shifts place the entire selected window outside the reference bands for Original and Shift ($F_{\min}>0$), and inside them for Widen and Joint ($G<0$). These sector assignments hold across all tested sizes and phases in Fig.~\ref{fig:inverse-design}(b).

The offsets used to design the third return also modify the first two windows. Both returns persist for Shift, Widen, and Joint, with the first window narrowing and the second widening relative to Original [Fig.~\ref{fig:all-return-design}]. The typical IPR dips follow the shifted complete-return intervals.

\bibliography{references}
\end{document}